\documentclass[aip, apl, reprint]{revtex4-1}
\pdfoutput=1 
\usepackage{graphicx}
\usepackage{dcolumn}
\usepackage[dvipsnames]{xcolor}
\usepackage{bm}
\usepackage{xr}
\usepackage{soul}
\usepackage[pagewise]{lineno}

\begin{document}

\title{Antisite defects stabilized by antiphase boundaries in YFeO$_3$ thin films}

\author{Abinash Kumar}
\affiliation{Department of Materials Science \& Engineering, Massachusetts Institute of Technology, Cambridge, MA 02139, USA}
\author{Konstantin Klyukin}
\affiliation{Department of Materials Science \& Engineering, Massachusetts Institute of Technology, Cambridge, MA 02139, USA}
\author{Shuai Ning}
\affiliation{Department of Materials Science \& Engineering, Massachusetts Institute of Technology, Cambridge, MA 02139, USA}
\author{Cigdem Ozsoy-Keskinbora}
\affiliation{Thermo Fisher Scientific, 5651GG Eindhoven, Netherlands}
\author{Mikhail Ovsyanko}
\affiliation{Thermo Fisher Scientific, 5651GG Eindhoven, Netherlands}
\author{Felix van Uden}
\affiliation{Thermo Fisher Scientific, 5651GG Eindhoven, Netherlands}
\author{Ruud Krijnen}
\affiliation{Thermo Fisher Scientific, 5651GG Eindhoven, Netherlands}
\author{Bilge Yildiz}
\affiliation{Department of Materials Science \& Engineering, Massachusetts Institute of Technology, Cambridge, MA 02139, USA}
\affiliation{Department of Nuclear Science \& Engineering, Massachusetts Institute of Technology, Cambridge, MA 02139, USA}
\author{Caroline A.~Ross}
\affiliation{Department of Materials Science \& Engineering, Massachusetts Institute of Technology, Cambridge, MA 02139, USA}
\author{James M.~LeBeau}
\affiliation{Department of Materials Science \& Engineering, Massachusetts Institute of Technology, Cambridge, MA 02139, USA}

\begin{abstract}

YFeO$_3$ thin films are a recent addition to the family of multiferroic orthoferrites where Y\textsubscript{Fe} antisite defects and strain have been shown to introduce polar displacements while retaining magnetic ordering. Complete control of the multiferroic properties, however, necessitates knowledge of the defects present and their potential role in modifying behavior. Here, we report the structure and chemistry of antiphase boundaries in multiferroic YFeO$_3$ thin films using aberration corrected scanning transmission electron microscopy combined with atomic resolution energy dispersive X-ray spectroscopy. We find that Fe\textsubscript{Y} antisites, which are not stable in the film bulk, periodically arrange along antiphase boundaries due to changes in the local environment. Using density functional theory, we show that the antiphase boundaries are polar and bi-stable, where the presence of Fe\textsubscript{Y} antisites significantly decreases the switching barrier. These results highlight how planar defects, such as antiphase boundaries, can stabilize point defects that would otherwise not be expected to form within the structure.

\end{abstract}

\maketitle

While the orthoferrite YFeO$_3$ (YFO) is centrosymmetric (non-polar) and antiferromagnetic in bulk \cite{Shang2013TheYFeO3,Shang2016TheYFeO3}, it exhibits multiferroic behavior when grown as a strained thin film under nominally stoichiometric and Y-rich conditions \cite{Ning2021}. Specifically, the introduction of Y\textsubscript{Fe} antisites breaks crystal inversion symmetry ($Pbnm$ to $R3c$) and stabilizes a spontaneous dipole moment that leads to stoichiometry-dependent ferroelectricity.  Although the resulting multiferroic behavior is robust, planar defects are also found in YFO thin films grown on SrTiO$_3$ \cite{Scola2011}, but the influence of these defects on behavior requires further investigation.

Planar defects, such as twins or antiphase boundaries (APBs), in the ferroic systems can modify both the local magnetic and polarization responses.  As a result, the behavior at the boundaries can differ greatly from the rest of the material. For example, twins in CaTiO$_3$ have been shown to exhibit polar displacements that lead to local ferrielectricity \cite{VanAert2012, Morikawa2021}. Similarly, the presence of APBs in Fe$_3$O$_4$ has been found to antiferromagnetically couple adjacent ferrimagnetic domains \cite{McKenna2014a}. By controlling planar defect density via thin film growth, either through substrate surface steps \cite{YoshioITOH1990,Jiang2002,Hensling2019}, cation off-stoichiometry \cite{Xu2016}, or lattice symmetry mismatch between the thin film and substrate \cite{Luysberg2009}, the film properties can thus be tuned \cite{Wang2018}. 

Because antisite point defects play a critical role in determining the multiferroic properties in YFO, determining their interactions with the planar defects can provide additional insights into the properties. For example, the local structure and chemistry of planar defects can lower the formation energy of point defects \cite{Aggarwal:1998wz,Antonelli:1999us,He:2003tn,Zhao2019} that may not occur in the `bulk' of the crystal. In YFO, for example, Fe\textsubscript{Y} antisites have a significantly higher formation energy\cite{Ning2021} even though Y$^{3+}$ and Fe$^{3+}$ are isovalent. This is due to their atomic radius mismatch ($r$\textsubscript{Y} = 106 pm and $r$\textsubscript{Fe} = 65 pm \cite{Bharadwaj2020}). The bonding environment at the planar defects in YFO, however, could enable such point defects to form and offer an additional `knob' of property control through point defect engineering \cite{Tuller:2011td}.

The interaction between point and planar defects requires being able to directly probe the atomic and chemical structure. While single substitutional dopants and antisites \cite{Voyles:2002vv,Chung.2008} have been observed with annular dark-field scanning transmission electron microscopy (STEM), their detection can be ambiguous depending on the atomic numbers of the host and defects \cite{Mittal:2011ud} and/or the static andthermal displacements \cite{Kim:2016uw}.  These limitations can be overcome by combining imaging and atomic resolution electron energy electron or energy dispersive X-ray spectroscopies where sensitivty to single-or-few atom point defects, i.e.~antisite defects, is now possible \cite{Cherns.2018,Jung2020,Huang.2012,Ning2021}, thus enabling new opportunities to  investigate point defect-planar defect interactions that may have been previously undetected.


In this Letter, we determine the structure of APBs in multiferroic Y-rich YFeO$_3$ (YFO) thin films using aberration corrected scanning transmission electron microscopy (STEM). Via atomic resolution energy dispersive X-ray spectroscopy (EDS) acquired with a large collection solid angle ($>$4 strad), we confirm that the APBs host  Fe$_\mathrm{Y}$ antisites, which are not found elsewhere in the film. We also show that significant relaxation occurs  at the APBs.  Using the structure observed from STEM, we apply density functional theory (DFT) to show that the observed APBs have a low formation energy and exhibit a bi-stable polar distortion.  We show that the Fe$_\mathrm{Y}$ antisite formation energy and polarization switching barrier are reduced by a factor of three at boundaries, leading to changes in the local properties.

The Y-rich multiferroic YFO thin films studied here are grown  by pulsed laser deposition on Nb doped SrTiO$_3$ and are approximately 30 nm thick. With a nominal misfit of -1.5\%, the films exhibit an epitaxial relationship of $[001]$\textsubscript{YFO} $\parallel$  $[001]$\textsubscript{STO} $(001)$\textsubscript{YFO} $\parallel$ $(001)$\textsubscript{STO}, where the YFO directions are given using the $Pbnm$ spacegroup setting. Compared with a simple cubic perovskite, the orthorhombic YFO unit cell lattice parameters are a=5.587\AA, b = 5.274\AA and c=7.5951\AA, i.e. rotated by $\pi/4$ and doubled along the $c$ axis. 

\begin{figure}[!ht]
    \centering
    \includegraphics[width = 3.3 in]{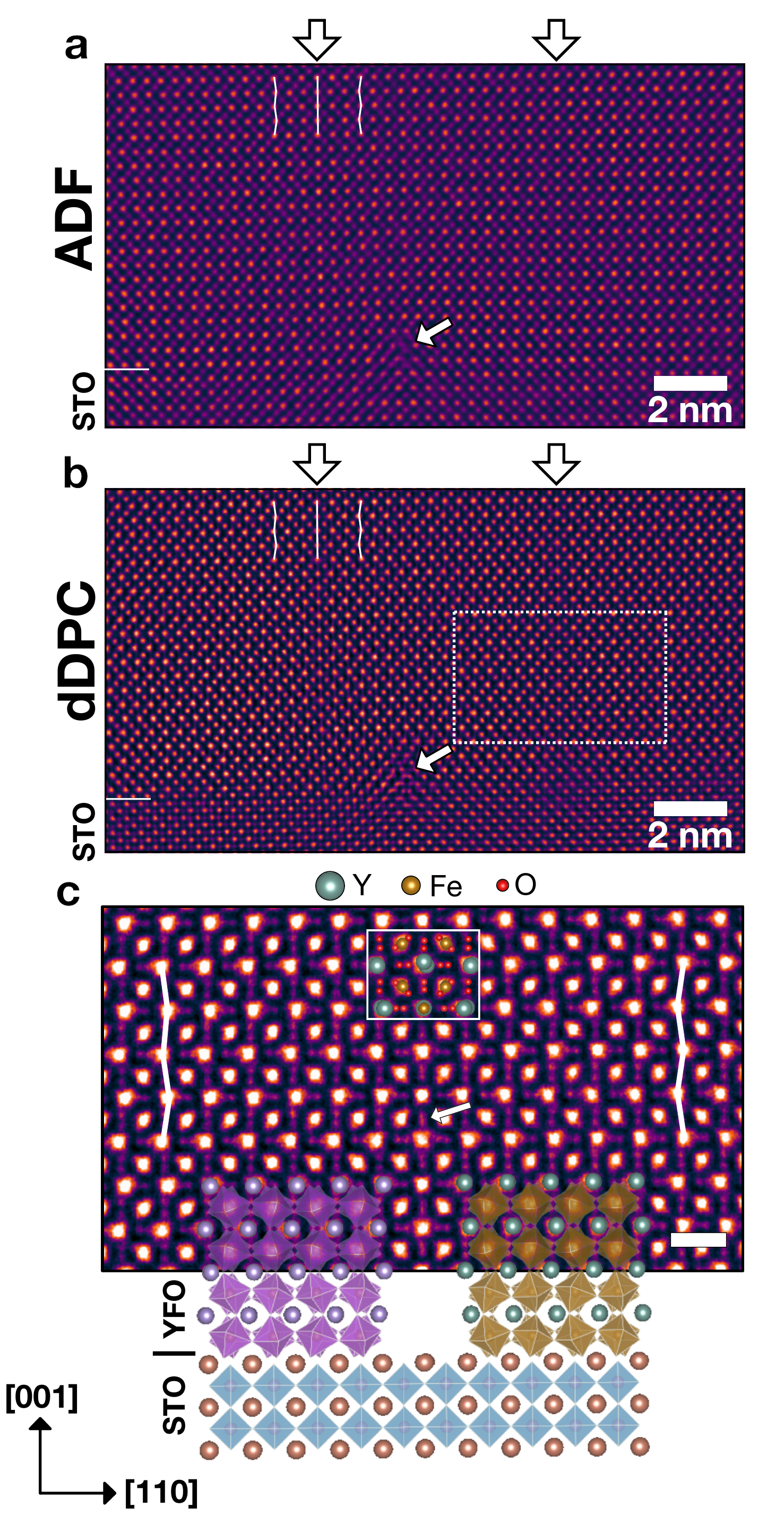}
    \caption{(a) ADF and (b) dDPC STEM images of the Y-rich YFO thin film grown on Nb:STO.  The horizontal lines at the left mark the film/substrate interface, while the arrows and chevrons indicate the positions of APBs. The arrows inside the figure indicate a misfit dislocation. (c) An atomic resolution dDPC image highlighting the cation and anion positions across the APB, where the inset provides the DFT relaxed structure.  The schematic shows that the $(110)$ APB forms as a result the initial growth plane [$(001)$, left or $(002)$, right] at the substrate. The arrow indicates positions where the projected O-O distance decreases at APB. The scale bar represents 500 pm.}
    \label{fig:APBs}
\end{figure}

Using aberration corrected annular dark-field (ADF) and differentiated differential phase contrast (dDPC) STEM, Figure \ref{fig:APBs}a,b, the substrate/film interface and defects can be identified.  The STO substrate can be distinguished by the slight decrease in intensity in the lower sixth of both images. Two APBs, indicated by arrows at the top of Figures \ref{fig:APBs}a,b, occur where the  zigzagging of the Y atom columns is mirrored on either side of the boundaries (see chevron lines in Figure \ref{fig:APBs}), and represents a translation of $\frac{c}{2}[001]$ in the $(110)$ plane. Further, the APBs are found throughout the thin film with a lateral density of about 0.2 APB/nm. While interface misfit dislocations are observed, marked with the arrows in Figures \ref{fig:APBs}a,b, they do not correlate with APB formation.

Unlike APBs in simple perovskites, such as SrTiO$_3$, the APB defects seen here do not arise from off-stoichiometry in the film. The boundaries are generated by a half-unit-cell translation along the doubled pseudocubic unit cell direction, $[001]$, and hence stoichiometry is maintained.  Furthermore, the APBs do not correlate with substrate surface steps and the YFO $d_{002}$ spacing is nearly that of STO, with a -2.7\% lattice  mismatch. Thus, an APB would not be required to accommodate a steps on the SrTiO$_3$ surface \cite{Wang2018}. Moreover, the APBs are found throughout the film occur at random positions. 

\begin{figure}[!ht]
    \centering
    \includegraphics[width = 3.3in]{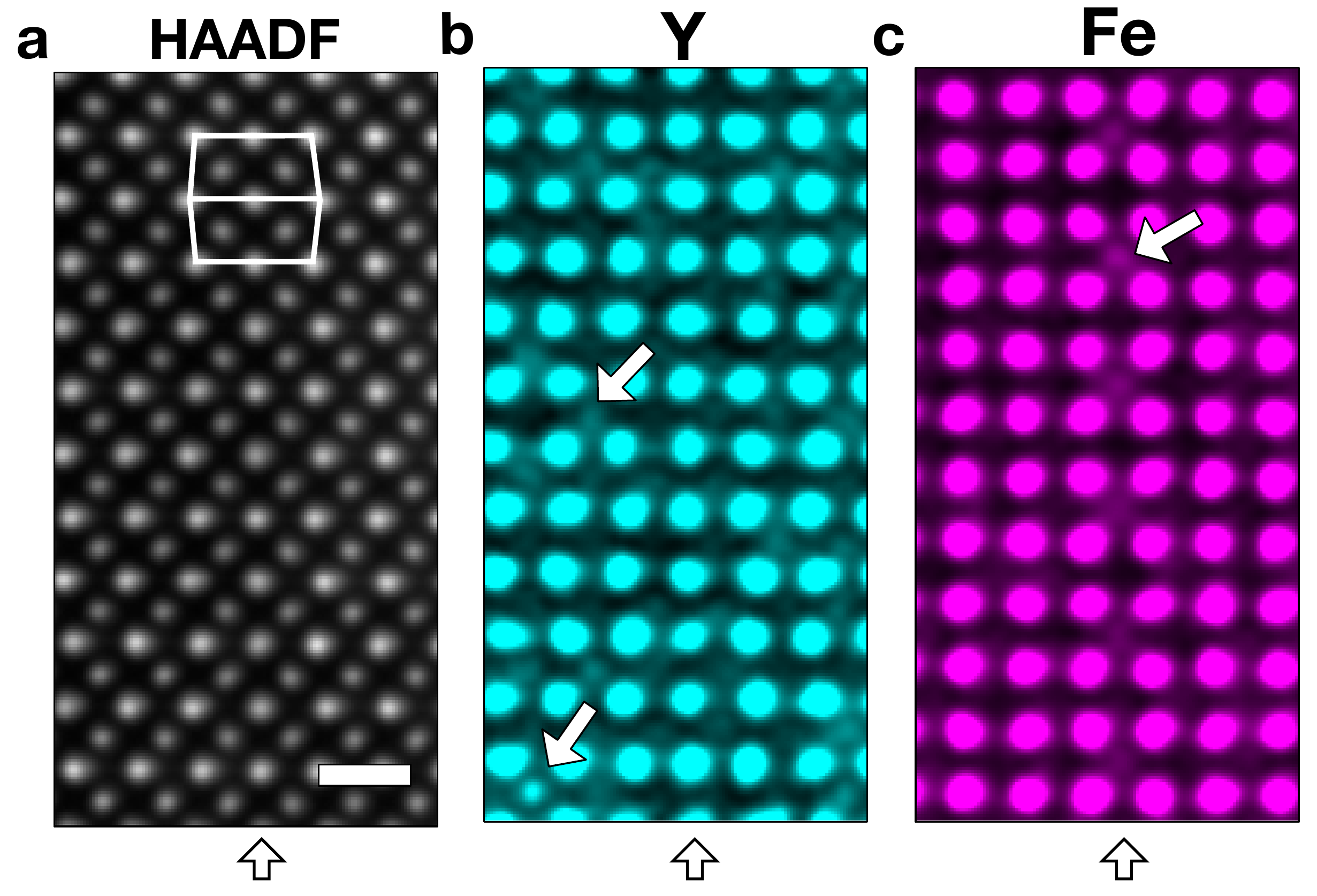}
    \caption{(a) HAADF STEM of an $(110)$ antiphase boundary in YFO and corresponding (b) Y and (c) Fe STEM EDS maps. The scalebar represents 500 pm. The arrows in (b) and (c) point to atom columns containing Y$_\mathrm{Fe}$  and Fe$_\mathrm{Y}$ antisites, respectively.}
    \label{fig:atomicEDS}
\end{figure}

As the orthorhombic YFO has lower symmetry than the cubic SrTiO$_3$ (STO) substrate, several equivalent nucleation planes are possible. Specifically, the substrate surface can equivalently accommodate nucleation on either $(001)$\textsubscript{YFO} or $(002)$\textsubscript{YFO}.  As shown in the schematic of Figure \ref{fig:APBs}c, these initial YFO configurations differ only by tilting of Fe-oxygen octahedra where the two stacking sequences form an APB when brought together. The observed APBs are thus the product of nucleation and growth during PLD. 
High-angle ADF (HAADF) and atomic resolution energy dispersive X-ray spectroscopy (EDS) confirm that the APBs are nominally composed of Y, as shown in Figures \ref{fig:atomicEDS}a,b and expected from the formation mechanism. Further, the Y EDS map shows that  Y\textsubscript{Fe} antisites form throughout the film, as previously reported in Ref.~\citenum{Ning2021}. These antisites have been identified as being responsible for the ferroelectric behavior measured in these Y-rich films.  The Fe EDS map, on the other hand, shows that  Fe\textsubscript{Y} reside at the APB, as in Figure \ref{fig:atomicEDS}c. Further, they are placed at every other Y atom column in the APB. This observation strongly suggests that while the formation energy of  Fe\textsubscript{Y} antisites is high in `bulk', it is significantly decreased at the APBs.  

\begin{figure*}[!ht]
    \centering
    \includegraphics[width = 6.5 in]{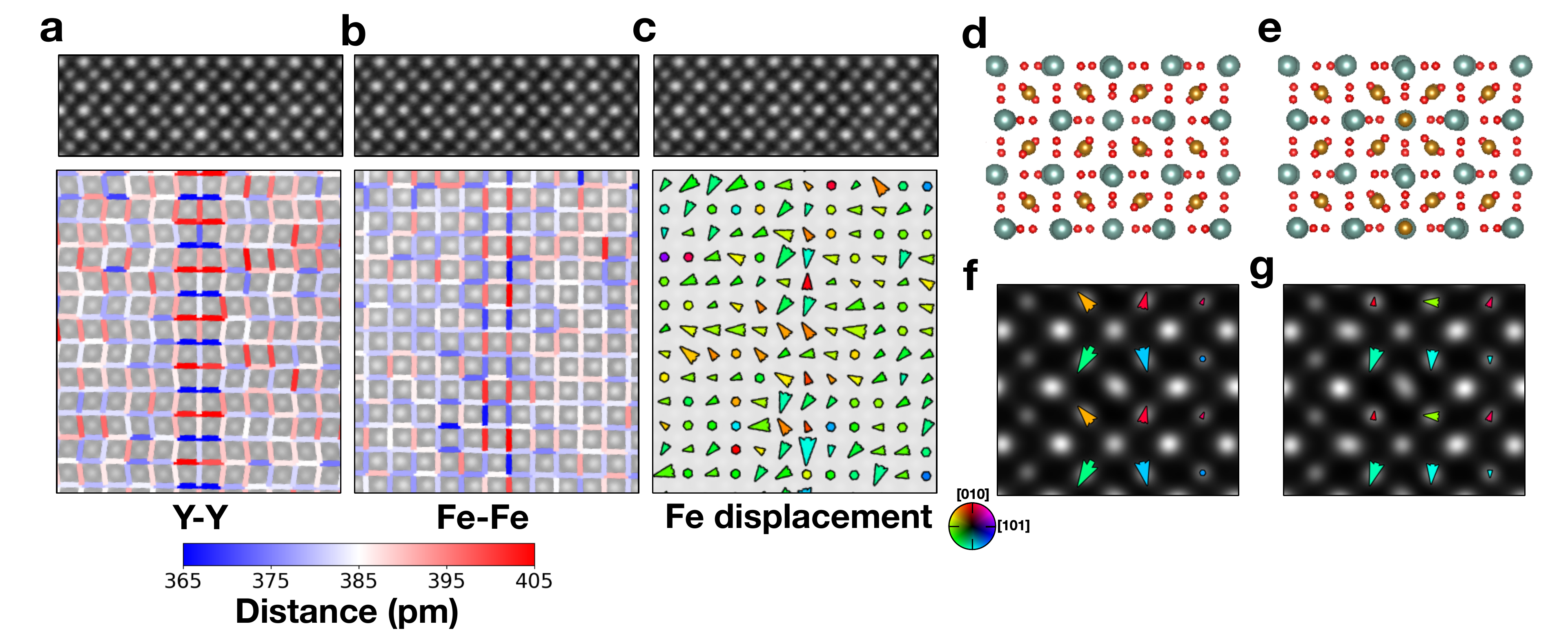}
    \caption{(a) Y-Y) and (b) Fe-Fe NLN distances from ADF STEM. (c) The net Fe displacement map obtained from the atom columns positions. (d, e) relaxed DFT structure of APB without and with an Fe$_\mathrm{Y}$ antisite respectively, (f, g) simulated ADF STEM image of APB  relaxed DFT structure  without and with an Fe$_\mathrm{Y}$ antisite,  respectively.}
    \label{fig:Struct}
\end{figure*}

To further explore the APBs, structural relaxations at the boundary are measured. The nearest-like-neighbor (NLN) distances for Y are shown in Figure \ref{fig:Struct}a.  Notably, the Y atom columns align vertically along $\left[001\right]$\textsubscript{YFO} at the APB in contrast to the zigzag to either side of the boundary. Furthermore, the in-plane Y-Y NLN distances alternate between expansion and contraction by 15\%  while the in-plane Fe-Fe NLN distances remain constant across the boundary.  In contrast, the out-of-plane NLN distances for Y-Y are unchanged and the Fe-Fe distances alternately expand and contract by 7\%. The oxygen atom positions also relax at the boundary, where along $[001]$ the projected oxygen positions move towards the Y atom column within the expanded lattice environment, as shown in Figure \ref{fig:APBs}c. These changes represent a significant departure from the `pristine' crystal structure, and hence the bonding environment is considerably different at the APB.

The net cation displacements at the APB, measured as the difference between the position of an Fe atom column and the  centroid of its four surrounding Y atom columns, reflect the departure from the bulk structure symmetry.  The average net cation displacement magnitude in Figure \ref{fig:Struct}c is 9$\pm$3 pm at the APB, and only 6$\pm$3 pm away from the boundary. The net cation displacements can thus be understood as inversion symmetry breaking that leads to polarization at the APB. From these measurements of the projected structure, the APBs exhibit ferrodistortive displacements, with the largest component of polarization along the in-plane direction.

\begin{figure}
    \centering
    \includegraphics[width=3.5 in]{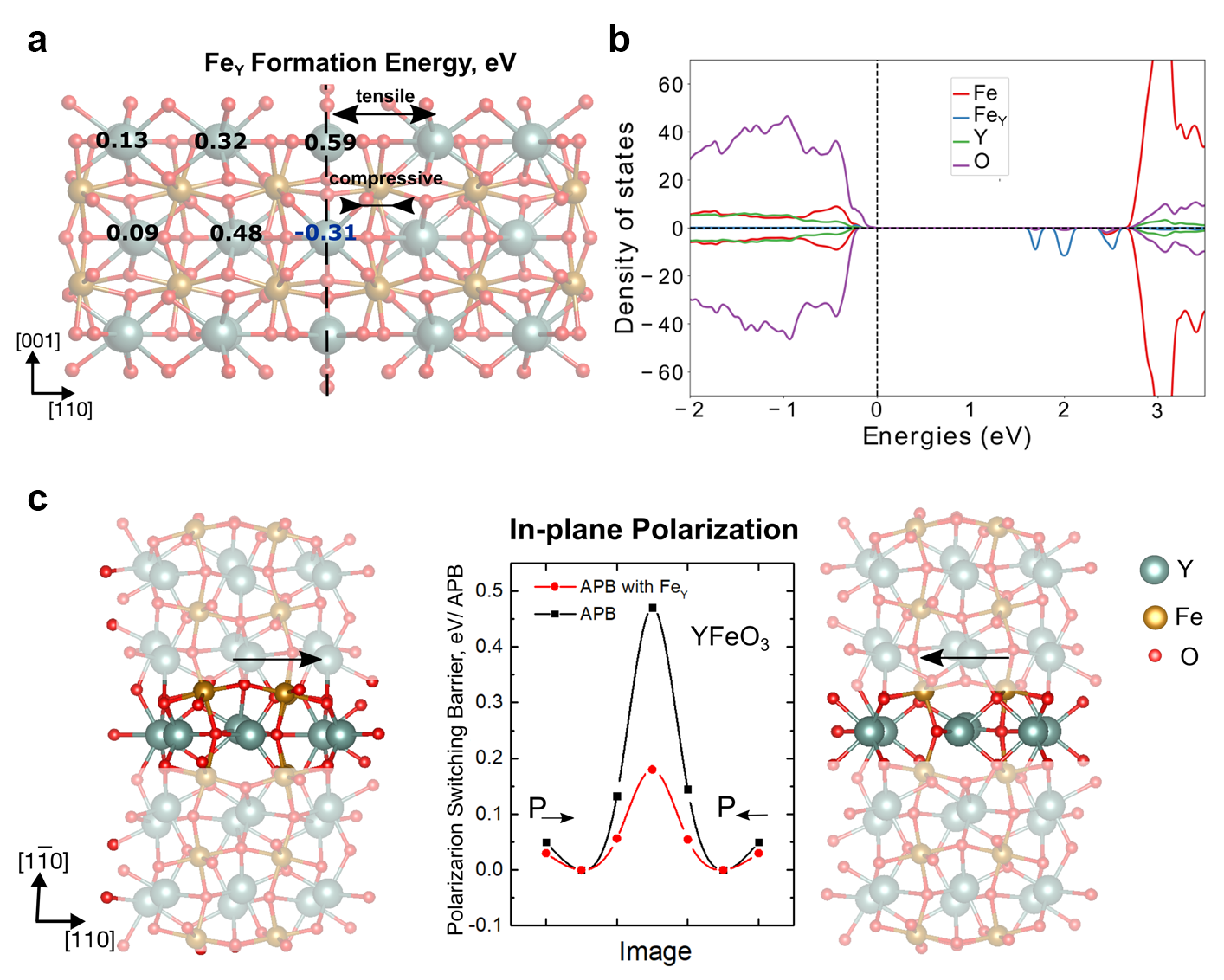}
    \caption{ (a) The relative formation energies of Fe\textsubscript{Y} defects (eV) at different Y positions across an APB.  A Fe\textsubscript{Y} in bulk YFO was used for the reference energy. (b) Projected density of states for Fe-rich APB of YFO where the Fermi level is set to zero.
    (c) DFT calculations demonstrating the  bi-stable, polar nature of the $(101)$ APB in YFO.}
    \label{fig:dft}
\end{figure}

The combination of polar displacements and the presence of Fe$_\mathrm{Y}$ antisites strongly suggests that the properties differ significantly at the APBs. Exploring this further, DFT is used to relax the structure of the APB measured from experiment, which was then used to estimate formation energies and local polarization with and without Fe antisites, Figures \ref{fig:dft}d,e. To confirm that the DFT APB structure agrees with experiment, ADF STEM images are simulated using relaxed DFT structures, Figures \ref{fig:dft}f,g.  The measured magnitude of the net cation displacement using the simulated APB images 6 pm with and 9 pm without the Fe$_\mathrm{Y}$ antisite, respectively. Furthermore, the Y and Fe sublattices at the APB of the DFT-relaxed structure also expand and contract by 16\% (along in-plane) and 8\% (along out-plane). The DFT structure is thus in excellent agreement with experiment.

The formation energy of the observed APBs is found to be 130 mJ/m$^2$ from DFT, which is within the range typical for perovskite oxides (100-300 mJ/m$^2$) \cite{hirel2010theoretical,Wei2014}. The low boundary formation energy further supports the nucleation and growth model suggested above as the boundaries can form without a large energy penalty.  Furthermore, the DFT calculations indicate that the Fe\textsubscript{Y} antisites are stabilized by the APBs where the antisite formation energy is decreased ($\Delta E = -0.31$ eV) at the compressive strain locations (see Figure \ref{fig:dft}a).  This is significant as at these Y positions, the Y-O distance is only 2.23 \AA{} compared to 2.29 \AA{} away from the boundary. Thus, the replacement of  Y by Fe is favored here because of Fe's smaller ionic radius, which reduces the compressive strain energy. This is in contrast to the Y APB positions with tensile strain, where a  significant increase ($\Delta E = 0.59$ eV) of the Fe$_\mathrm{Y}$ formation energy occurs. Furthermore, the DFT calculations also show that while the electronic structure of Fe-rich APB remains insulating,  empty states associated with the Fe$_\mathrm{Y}$ defects (see Figure \ref{fig:dft}b) may serve as charge traps and promote p-type conductivity.

As with experiment, in-plane polarization induced by the APBs is also found in the first-principles simulations. The DFT structure shows that in-plane polarization originates from significant off-centering (0.71 \AA) of central Y atom at the APB as shown in Figure \ref{fig:dft}c, which polarizes adjacent layers and causes sizable octahedral distortions. The resulting bi-stable polarization is 19 $\mu$C/cm$^2$ along $[110]$\textsubscript{YFO} with a switching barrier of 0.44 eV/APB. Moreover, bi-stable APBs have also been shown to occur in other perovskite oxides \cite{Wei2014}. When Fe\textsubscript{Y} antisites are added to the APBs, the off-centering is reduced by more than a factor of three, 0.2 \AA, which leads to a decreased ferroelectric polarization of 7.6 $\mu$C/cm$^2$ (Figure \ref{fig:dft}c).  The smaller displacement along $[110]$\textsubscript{YFO} in the DFT relaxed structure is associated with the smaller ionic radius of Fe$^{3+}$ compared to Y$^{3+}$  and ultimately reduces the switching barrier to  0.15 eV/APB. This is significant as a boundary without Fe\textsubscript{Y} antisites would exhibit a large switching barrier and could thus act to pin polarization, which can degrade  switching behavior. The decrease in APB polarization with Fe\textsubscript{Y} antisites, along with the measured, robust response of the multiferroic YFO films \cite{Ning2021}, thus suggests that the antisite doped APBs do not significantly affect the saturation polarization. 

Beyond the polarization, magnetic properties can be impacted by presence of APBs. In the DFT relaxed structure, the oxygen octahedra at the APBs are found to tilt. This can be observed when viewed normal to $(1\bar{1}0)$\textsubscript{YFO}, where the separation between the oxygen atom columns decreases at the boundary where tensile strain occurs (inset Figure \ref{fig:APBs}c), which is also observed in experiment as marked by arrow in Figure \ref{fig:APBs}c. Measured from the DFT structure, the angle between two octahedra along $\left[110\right]$ is 140$^\circ$ in centrosymmetric YFO while at the boundary this angle alternates between 139$^\circ$ and 156$^\circ$. Moreover, the oxygen octahedra at the boundary distort along $\left[001\right]$ where the Fe-O distances become asymmetric with a ratio of 1.12. As a consequence of these Fe-O bond angles and distances at the APB, the magnetic properties of the boundary are likely to differ from the rest of the film. For example, superexchange coupling between Fe and O depends on the Fe-O bond distance and angle, which can lead to antiferromagnetic behavior, as seen in  SmFe$_x$Cr$_{1-x}$O$_3$ \cite{Xiang2018}. Thus, these observations point to the need for additional local magnetic property studies of the APBs.  

In summary, the presence of APBs in multiferroic YFO thin films can stabilize Fe$_\mathrm{Y}$ antisites that are otherwise unfavored in bulk. Through direct, atomically-resolved imaging, the APBs exhibit significant structural relaxation of the Y, Fe and O sub-lattices at the boundary. The combination of these STEM measurements with DFT calculations show that the APBs provide a local structural and chemical environment that lowers the formation energy of Fe$_\mathrm{Y}$ antisites considerably. The local distortions at the APBs are also shown to be  ferrodistortive in nature, which is be modified by the presence of Fe$_\mathrm{Y}$ defects. Specifically, the bi-stable switching barrier is reduced by a factor of about three, which would reduce or eliminate potential polarization pinning.  The results thus indicate that APBs can provide an additional means to control the multiferroic properties of orthoferrites.  Finally, we suggest that the mechanism for point defect stabilization by APBs should be common in other functional oxides where the pseudocubic unit cell is doubled along one or more of the crystal axe, which may provide a means to locally introduce dopants that would otherwise be unstable.


\subsection*{Acknowledgements}

JML and AK acknowledge support of this work through the John Chipman Career Development Professorship. AK thanks the MIT Mathworks engineering fellowship for support. The DFT calculations were carried out using the Extreme Science and Engineering Discovery Environment (XSEDE) \cite{towns2014xsede}, which is supported by National Science Foundation Grant No. ACI1548562. The thin film growth was supported by the MRSEC Program of the National Science Foundation under award No.~DMR-1419807. The NVIDIA Titan Xp GPU used for this research was donated by the NVIDIA Corporation.  This work was carried out in part through the use of the MIT Characterization.nano facility.


\bibliography{refs}

\section*{Methods}

\subsection*{Thin film growth}
The YFO thin film was grown using pulsed laser deposition on an Nb-doped STO substrate using a KrF excimer laser ($\lambda$ = 248 nm) with 1.3 J/cm$^2$ fluence and 10 Hz of repetition rate \cite{Ning2021}. A commercial YFeO$_3$ target was used for thin film growth. The substrate was held at 900 $^\circ$C and the oxygen partial pressure was kept at 10 mTorr. The as-grown thin films were cooled to room temperature under a similar partial pressure of oxygen with the rate of 20 $^\circ$C/min. 

\subsection*{Scanning Transmission Electron Microscopy}

Cross-sectional samples of YFO thin films were prepared for electron microscopy using conventional polishing using an Allied Mulitprep system.  A Fischione 1051 argon ion mill was used to thin the samples to electron transparency.  Scanning transmission electron microscopy imaging was conducted using a probe-aberration corrected Thermo Fisher Scientific Themis Z G3 S/TEM 60-300kV equipped with an XFEG source operated at 200 kV. The STEM images were acquired with a convergence semi-angle of 18 mrad (ADF) or 25 mrad (dDPC).  The images used for structural analysis were acquired using the revolving STEM (RevSTEM) method to ensure image accuracy and precision \cite{Sang2014c, Dycus2015}. Each RevSTEM dataset  consisted  of  20 frames  with the fast scan direction rotated 90$^\circ$ between each. The atom column locations were extracted from the drift and scan distortion corrected images using a custom Python script \cite{Kumar2021}.

Atomic resolution EDS data was collected using an Ultra-X detector (>4 strad collection solid angle) equipped on a Thermo Fisher Scientific Spectra Ultra microscope with a X-CFEG source operated at 200 kV. The probe convergence semi-angle was 18 mrad and the electron dose was 4.35x10$^3$ e$^-$/pixel. The atomic resolution EDS dataset was processed using non-local principle components analysis to reduce noise and Gaussian blurring via an open-source Matlab script \cite{Salmon2014}.  STEM image simulations were carried out using the multislice approach \cite{Kirkland2010} with imaging conditions from experiment.  The simulated sample thickness was 10 nm to match that from experiment determined using position averaged convergent beam electron diffraction \cite{Lebeau2010}. To approximately account for the finite effective source size, simulated images were convolved with an 80 pm full-width at half-maximum Gaussian \cite{Lebeau2008}.

\subsection*{Density Functional Theory and Image Simulations}

First-principles calculations were performed within  density functional theory (DFT)  using the projector augmented wave (PAW) potentials \cite{blochl1994projector} as implemented in the Vienna Ab initio Simulation Package (VASP)\cite{kresse1996efficiency}. The generalized gradient approximation Perdew-Burke-Ernzerhof (PBE) exchange-correlation functional \cite{perdew1996generalized} was employed with a plane wave cutoff energy of 500 eV. The rotationally invariant PBE + U approach was adopted with U\textsubscript{eff} = 4 eV on the Fe 3d orbitals. The ions were relaxed by applying a conjugate-gradient algorithm until the Hellmann-Feynman forces were less than 10 meV/\AA. In-plane lattice parameters were fixed to simulate epitaxial growth on a cubic SrTiO$_3$ substrate (a=3.903 \AA).  Ferroelectric properties were calculated using the Berry-phase approach \cite{king1993theory}. Switching barriers were calculated using Nudged Elastic Band method and electronic structure analysis was carried out using the HSE06 functional\cite{krukau2006influence}. Anti-ferromagnetic G-type spin-ordering was imposed at the APB. 

To implement periodic boundary conditions, a large supercell was considered with two antiphase boundaries along $\left[110\right]$ of the $Pbnm$ structure.  The resulting supercell was comprised of 8 pseudocubic ABO$_3$ unit cells along a direction, $y$, perpendicular to the APB and 2 unit cells along directions $x$ and $z$ parallel to the APB. The APB energy was computed as $E$\textsubscript{APB} = $(E $-$ E_0)/2S$, where E is the total energy of the APB configuration, E$_0$ the energy of the single-domain supercell of the same size and $S$ is equal to the cross-sectional area of the supercell.

\end{document}


\maketitle

\tableofcontents
\listoffigures

\baselineskip24pt



\section{Experimental Section}
\label{suppl:Experiment}

\section{APB contrast in ADF and iDPC image}
\label{suppl:image contrast}

\printbibliography

\newpage